# Empirical Evaluations of Seed Set Selection Strategies for Predictive Coding

Christian J. Mahoney
e-Discovery
Cleary Gottlieb Steen & Hamilton LLP
Washington, D.C. USA
Email: cmahoney@cgsh.com

Nathaniel Huber-Fliflet
Data & Technology
Ankura Consulting Group, LLC
Washington, D.C. USA
Email: nathaniel.huber-fliflet@ankura.com

Katie Jensen
Data & Technology
Ankura Consulting Group, LLC
New York, NY USA
Email: katie.jensen@ankura.com

Dr. Haozhen Zhao
Data & Technology
Ankura Consulting Group, LLC
Washington, D.C. USA
Email: haozhen.zhao@ankura.com

Robert Neary
Data & Technology
Ankura Consulting Group, LLC
Washington, D.C. USA
Email: robert.neary@ankura.com

Shi Ye
Data & Technology
Ankura Consulting Group, LLC
Washington, D.C. USA
Email: shi.ye@ankura.com

*ABSTRACT– Training documents have a significant impact on the performance of predictive models in the legal domain. Yet, there is limited research that explores the effectiveness of the training document selection strategy – in particular, the strategy used to select the seed set, or the set of documents an attorney reviews first to establish an initial model. Since there is limited research on this important component of predictive coding, the authors of this paper set out to identify strategies that consistently perform well. Our research demonstrated that the seed set selection strategy can have a significant impact on the precision of a predictive model. Enabling attorneys with the results of this study will allow them to initiate the most effective predictive modeling process to comb through the terabytes of data typically present in modern litigation. This study used documents from four actual legal cases to evaluate eight different seed set selection strategies. Attorneys can use the results contained within this paper to enhance their approach to predictive coding.*

## I. INTRODUCTION

"In 2018, the total number of business and consumer emails sent and received per day exceeded 281 billion and is forecast to grow to over 333 billion by year-end 2022" [1]. This rapid data growth presents cost and resource challenges to corporations dealing with internal investigations and legal or regulatory actions. Corporations are required to produce relevant data to the requesting party when responding to litigation or an enforcement agency inquiry. The process used to respond to legal matters involves attorneys manually reviewing and classifying large volumes of documents into categories, such as relevant or not relevant to the matter and to production requests from parties or regulators. This requirement presents corporations, and their legal teams, with the prospect of collecting and reviewing millions of documents, at great cost and under tight deadlines, to determine what is relevant to the matter at hand. Legal document review can consume over 70 percent of a litigation's discovery budget [2]. Legal teams employ many strategies to cull these vast data sets into manageable review populations to lower costs.

Traditional culling methods, such as keyword searching, have had difficulty keeping up with the rapid growth of data volumes. In practice, keywords tend to be broad and return false positives, often leading to review populations that are still too cumbersome to cost-effectively review. As data sets continue to grow, it is no surprise that the application of text classification, referred to as Predictive Coding or Technology Assisted Review (TAR) in the legal domain, has been on the rise for the last ten years. Norton Rose Fulbright's 2017 Litigation Trends Survey found 64 percent of respondents used TAR in the last 12 months [3]. This is up from 60 percent as reported in their survey from 2016, and 57 percent from 2015 [4].

The seed set is the initial set of documents used to train the machine and it is a critical component of the predictive model – it is the foundation on which a predictive model is built. A machine learning algorithm uses the labeled documents in the seed set to create the first iteration of a predictive coding enabled document review. This initial predictive model is typically used to prioritize the most relevant documents for review and initiate the document review process. This first iteration of the predictive coding process has a significant impact in the future direction of the document review and its overall cost.

Creating a performant predictive model with the fewest number of documents has benefits that include, reducing the time and expense required to train the model and iterate the predictive coding process, creating a model that meets the desired level of performance, and beginning the document review or producing documents to the requesting party faster. The size of the seed set effects how quickly a predictive model can be created because the attorney review of these documents generally takes longer to perform than the time for the machine to generate the model.

Not only does the number of seed set documents impact the effectiveness of the predictive coding process, but so does the content of the seed set documents. Seed sets that misrepresent the data set or contain mislabeled documents can impact model performance and cause unnecessarily low precision or recall. Predictive models with weak performance have a negative impact on the document review process and lead to issues, such as:

- Postponing the start of the review – deadlines are often tight and losing days or hours can delay attorney case factfinding, increase the cost to complete the review, and jeopardize completing the review on time.

- Reducing the model's precision and require the review of unnecessary documents – attorney review of irrelevant documents increases the time to complete the review and, in turn, the cost of the review.

- Requiring additional training review by senior attorneys – more training review can divert senior attorneys away from managing case strategy and other important client legal needs.

- Missing quality control opportunities – legal teams often compare document labels applied by attorney reviewers against the predictive model to identify documents that are potentially mislabeled. Consider a document that is labeled irrelevant but received a score that indicates it is relevant. This document would be selected for rereview, but the probability score needs to be reliable and if not, could jeopardize the accuracy of quality control workflows.

While we know that the seed set has a significant impact on the quality a predictive model, research is limited that explores the most effective seed set selection strategies. Attorneys are often left making an educated guess as to what strategy might work best for their legal matter. As a result, with the exception of a small number of technologists, the legal community has viewed the strategy to select seed set documents as more of an art than a science.

To date, legal industry standard practices recommend two widely-used selection strategies: (i) random sampling and (ii) judgmental sampling (e.g. keyword searching) [7][8][9].

- Random selection on a large population is diverse and allows for avoiding bias in selection but with low richness populations is not balanced enough [7][8][9].

- Judgmental selection can provide relevant features but introduces bias and studies have shown that keywords are unreliable [8][9].

The goal of our research was to identify seed set selection strategies that minimize seed set size, increase recall and precision, and supplement this topic's existing research. We evaluated the effectiveness of eight seed set selection strategies to provide attorneys with scientific analysis that they can use to influence their predictive coding process in the future. In this paper, we (i) describe the data sets and experiments; and (ii) detail the results of our findings, highlighting the seed set selection strategies that have the most positive impact on results.

## II. DATA SETS AND EXPERIMENT DESIGN

In this section, we describe the data sets and experiments. The experiments were designed to evaluate the impact of seed set document selection strategies on the initial round of the predictive coding process.

### A. DATA SETS

We conducted experiments on four data sets from confidential, non-public, actual legal matters across various industries such as telecommunications, construction, and consumer services. Each data set contained email, Microsoft Office documents, and other text-type documents. The four data sets ranged from 277,745 to 412,880 documents with each set's richness (or positive class rate) between 3.63 percent and 38.58 percent.

Attorneys labeled each data set's documents over the course of the legal matter and their review-coding labels provided the ability to evaluate the performance of the models. Additionally, attorneys created and used keyword lists to target responsive or privileged documents during these matters and those lists were available for our study. The data sets represent real-world scenarios and provide a feature rich environment with which to perform our experiments. They are compelling since they are diverse in content and have labels for responsiveness and privilege.

Tables 1A, 1B, 1C, and 1D detail the document populations of each data set: the total documents, review coding label, richness, and number of keyword terms.

Table 1A: Privilege Data Set Statistics

| Data Sets | Total Documents | Privileged Documents | Not Privileged Documents | Richness |
|---|---|---|---|---|
| Project A | 308,621 | 46,730 | 261,891 | 15.14% |
| Project B | 393,745 | 14,307 | 379,438 | 3.63% |
| Project C | 277,745 | 38,834 | 238,578 | 14.00% |

Table 1B: Responsive Data Set Statistics

| Data Sets | Total Documents | Responsive Documents | Not Responsive Documents | Richness |
|---|---|---|---|---|
| Project D | 412,880 | 159,304 | 253,576 | 38.58% |

Table 1C: Privilege Keyword Statistics

| Data Sets | Total Documents | Keywords | Documents Containing Keyword Hits | Keyword Hit Percentage |
|---|---|---|---|---|
| Project A | 308,621 | 808 | 193,017 | 62.54% |
| Project B | 393,745 | 4,211 | 368,506 | 93.59% |
| Project C | 277,745 | 509 | 159,900 | 57.57% |

Table 1D: Responsive Keyword Statistics

| Data Sets | Total Documents | Keywords | Documents Containing Keyword Hits | Keyword Hit Percentage |
|---|---|---|---|---|
| Project D | 412,880 | 23 | 81,362 | 19.71% |

**B. EXPERIMENT DESIGN**

Each data set was divided into a selection population and a testing population. The testing population consisted of ten percent of the overall data set – selected at random – and, once selected, remained static during all experiments. The seed set documents were selected from the remaining 90 percent of the data set.

Three sizes of seed sets were created – 500 documents, 1,000 documents, and 2,000 documents – and eight seed set selection strategies were applied to each seed set size resulting in 24 predictive models per project. The predictive model text preprocessing parameters and machine learning algorithm remained constant for each experiment. The experiments used a bag of words approach and Table 2 highlights key modeling parameters.

Table 2: Predictive Model Parameters

| Parameters | Parameter Value |
|---|---|
| Word Stemming | No |
| N-Grams | 1 |
| Token Value Type | Normalized Frequency |
| Number of Tokens | 20,000 |
| Machine Learning Algorithm | Logistic Regression |

**Selection Strategies**

Eight selection strategies were used to identify seed set documents and the following is a detailed description of each.

1. Random Sample

Generate a random sample of documents pulled from the selection population for each seed set size (500, 1,000, 2,000).

2. Stratified Keyword

Using the selection population,

   A. Generate an index for each keyword that lists all documents which contain the keyword.
   B. Select at least one document from each keyword index.
   C. Ensure the same document was not selected from more than one keyword index.
   D. When the total number of selected documents exceeds the seed set size, the seed set was randomly selected from all the available keyword documents in the selected population.

Create a seed set for each seed set size (500, 1,000, 2,000).

3. Stratified Keyword – Weighted

Using the selection population,

   A. Generate an index for each keyword that lists all documents which contain the keyword.
   B. Calculate the percentage of the documents from each keyword index in relation to the overall Keyword population.
   C. Randomly select a number of sample documents from each keyword index commensurate to each index's percentage of the overall Keyword population.
   D. Ensure the same document was not selected from more than one keyword index.
   E. When the total number of available documents exceeds the seed set size, the seed set was randomly selected from all the available keyword documents in the selection population.

Create a seed set for each seed set size (500, 1,000, 2,000).

4. Keyword Model – Top Scoring

Using the selection population,

   A. Combine all keywords to form a linear model in which all keywords receive the same weight.
   B. Generate an index for each keyword that lists all documents which contain the keyword.
   C. Assign one point for each keyword hit per document.
   D. Rank the population according to the number of points assigned per document, highest to lowest.
   E. Select sample documents in order of points.

Create a seed set for each seed set size (500, 1,000, 2,000).

5. Keyword Model – Stratified

Using the selection population,

   A. Combine all keywords to form a linear model in which all keywords receive the same weight.
   B. Generate an index for each keyword that lists all documents which contain the keyword.
   C. Assign one point for each keyword hit per document.
   D. Rank population according to the number of points assigned per document, highest to lowest.
   E. Divide population into ten groups based upon points scored, each group contains ten percent of the population.
   F. Select an equal number of sample documents from each of the ten groups.

Create a seed set for each seed set size (500, 1,000, 2,000).

6. Weighted Keyword Model – Stratified

Using the selection population,

   A. Combine all keywords to form a linear model in which all keywords receive the same weight.
   B. Generate an index for each keyword that lists all documents which contain the keyword.
   C. Assign one point for each keyword hit per document.
   D. Rank population according to the number of points assigned per document, highest to lowest.
   E. Divide population into ten groups based upon points scored, each group containing ten percent of the population.
   F. Randomly select a number of sample documents from each group commensurate to each group's percentage of the overall Keyword population.

Create a seed set for each seed set size (500, 1,000, 2,000).

7. Clustering

Using the selection population,

   A. Generate 243 clusters using a variant of the K-Means clustering algorithm to create a cluster set of three branches to a depth of five layers.
   B. Randomly select an equal number of sample documents from each cluster.
   C. When the total number of available documents exceeds the seed set size, the seed set was randomly selected from all the available documents in the selection population.

Create a seed set for each seed set size (500, 1,000, 2,000).

8. Clustering – Weighted

Using the selection population,

A. Generate 243 clusters using a variant of the K-Means clustering algorithm to create a cluster set of three branches to a depth of five layers.
B. Calculate the percentage of the documents from each cluster in relation to the overall population.
C. Randomly select sample documents from each cluster commensurate to each cluster's percentage of the overall cluster population.
D. When the total number of available documents exceeds the seed set size, the seed set was randomly selected from all the available documents in the selection population.

Create a seed set for each seed set size (500, 1,000, 2,000).

**Experiment Implementation**

Each seed set selection strategy was used to create one seed set per seed set size across each project. The resulting seed set documents were used to create a predictive model and that model was applied to the corresponding testing population of each project. The performance of the model was then measured by examining precision at three specific recall levels: 50 percent, 75 percent, and 90 percent.

Table 3 broadly outlines the steps in each experiment.

Table 3: Experiment Procedure

| | |
|---|---|
| 1. | Compile seed set. |
| 2. | Train a model. |
| 3. | Score all the documents in the testing populations using the model. |
| 4. | Evaluate the performance of the model using each project's testing population and classifications. |

### III. RESULTS

In this section, we report and discuss the results of the experiments performed on Projects A, B, C, and D.

**A. Project A**

Tables 4A, 4B, and 4C illustrate the precision achieved at recall levels of 50 percent, 75 percent, and 90 percent using each strategy's seed set.

Table 4A: Project A - Seed Set 500 - Precision Rates at Recall Levels

| Seed Method | 90% | 75% | 50% |
|---|---|---|---|
| Stratified Keyword | 26.68% | 46.78% | 71.24% |
| Stratified Keyword – Weighted | 25.40% | 46.61% | 74.32% |
| Keyword Model – Stratified | 26.46% | 45.27% | 73.98% |
| Clustering | 24.26% | 44.69% | 70.02% |
| Random Sample | 23.62% | 43.08% | 65.18% |
| Weighted Keyword Model – Stratified | 23.26% | 42.13% | 66.89% |
| Clustering – Weighted | 24.92% | 41.69% | 66.79% |
| Keyword Model – Top Scoring | 19.05% | 22.30% | 27.36% |

Table 4B: Project A - Seed Set 1,000 - Precision Rates at Recall Levels

| Seed Method | 90% | 75% | 50% |
|---|---|---|---|
| Clustering – Weighted | 31.46% | 53.47% | 72.73% |
| Stratified Keyword – Weighted | 27.81% | 53.38% | 77.14% |
| Stratified Keyword | 28.93% | 53.27% | 78.89% |
| Keyword Model – Stratified | 28.20% | 52.98% | 76.33% |
| Weighted Keyword Model – Stratified | 29.43% | 52.26% | 76.16% |
| Clustering | 28.86% | 52.20% | 75.63% |
| Random Sample | 28.61% | 50.67% | 77.17% |
| Keyword Model – Top Scoring | 19.45% | 25.73% | 34.05% |

Table 4C: Project A - Seed Set 2,000 - Precision Rates at Recall Levels

| Seed Method | 90% | 75% | 50% |
|---|---|---|---|
| Random Sample | 36.32% | 59.82% | 79.97% |
| Stratified Keyword | 34.23% | 59.58% | 82.30% |
| Clustering | 37.44% | 59.13% | 79.05% |
| Keyword Model – Stratified | 32.60% | 56.96% | 76.91% |
| Stratified Keyword – Weighted | 35.29% | 56.87% | 81.11% |
| Weighted Keyword Model – Stratified | 33.51% | 56.79% | 78.05% |
| Clustering – Weighted | 34.80% | 55.89% | 77.42% |
| Keyword Model – Top Scoring | 21.52% | 28.00% | 36.51% |

The *Stratified Keyword* strategy achieved the highest precision at 75 percent recall using a seed set size of 500 documents, achieving 46.78 percent precision. The *Stratified Keyword – Weighted* strategy followed with 46.71 percent precision at the same recall level.

The *Clustering – Weighted* strategy achieved the highest precision at 75 percent recall using a seed set size of 1,000 documents, achieving 53.47 percent precision. The *Stratified Keyword – Weighted* strategy followed with 53.38 percent precision at the same level.

The *Random Sample* strategy achieved 59.82 percent precision at 75 percent recall using a seed set size of 2,000 documents and the *Stratified Keyword* strategy followed closely behind with 59.58 percent precision at the same recall level.

The *Random Sample* strategy, using a seed set size of 2,000 documents, achieved the highest precision of all the strategies at 75 percent recall. Further, predictive models created using seed set sizes of 2,000 documents achieved the top seven precision levels at 75 percent recall.

The Figure 1 illustrates the comparison of precision and recall levels on a curve for each strategy's seed set.

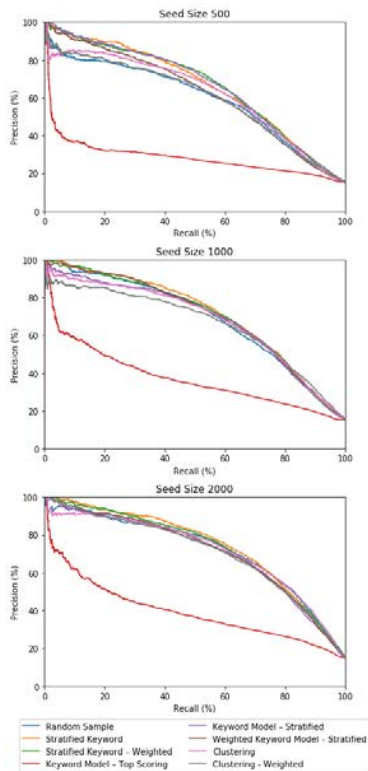

Figure 1. Project A - Precision & Recall by Strategy

## B. Project B

Tables 5A, 5B, and 5C illustrate the precision achieved at recall levels of 50 percent, 75 percent, and 90 percent using each strategy's seed set.

Table 5A: Project B - Seed Set 500 - Precision Rates at Recall Levels

| Seed Method | 90% | 75% | 50% |
|---|---|---|---|
| Stratified Keyword – Weighted | 4.85% | 6.02% | 8.56% |
| Stratified Keyword | 4.64% | 5.70% | 7.72% |
| Weighted Keyword Model – Stratified | 4.53% | 5.50% | 7.77% |
| Keyword Model – Stratified | 4.41% | 5.05% | 6.30% |
| Clustering | 4.22% | 4.72% | 6.26% |
| Clustering – Weighted | 4.32% | 4.71% | 5.18% |
| Random Sample | 4.22% | 4.59% | 5.30% |
| Keyword Model – Top Scoring | 4.08% | 4.26% | 4.28% |

Table 5B: Project B - Seed Set 1,000 - Precision Rates at Recall Levels

| Seed Method | 90% | 75% | 50% |
|---|---|---|---|
| Stratified Keyword – Weighted | 4.81% | 6.54% | 9.73% |
| Stratified Keyword | 4.79% | 6.35% | 9.86% |
| Keyword Model – Stratified | 4.96% | 6.34% | 8.56% |
| Clustering | 4.49% | 5.87% | 8.64% |
| Weighted Keyword Model – Stratified | 4.63% | 5.71% | 8.03% |
| Random Sample | 4.28% | 5.48% | 7.96% |
| Clustering – Weighted | 4.46% | 5.35% | 8.18% |
| Keyword Model – Top Scoring | 4.49% | 5.04% | 5.61% |

Table 5C: Project B - Seed Set 2,000 - Precision Rates at Recall Levels

| Seed Method | 90% | 75% | 50% |
|---|---|---|---|
| Random Sample | 5.26% | 7.30% | 10.54% |
| Clustering | 5.24% | 7.20% | 11.73% |
| Weighted Keyword Model – Stratified | 5.27% | 7.08% | 10.61% |
| Stratified Keyword | 5.18% | 7.05% | 11.07% |
| Stratified Keyword – Weighted | 5.13% | 7.05% | 12.16% |
| Keyword Model – Stratified | 4.98% | 6.75% | 9.72% |
| Clustering – Weighted | 4.53% | 6.45% | 10.54% |
| Keyword Model – Top Scoring | 4.88% | 5.55% | 6.87% |

The *Stratified Keyword – Weighted* strategy achieved the highest precision at 75 recall using a seed set size of 500 documents, achieving 6.02 percent precision. The *Stratified Keyword* strategy followed with 5.70 percent precision at the same recall level.

The *Clustering* strategy achieved the highest precision at 75 percent recall using a seed set size of 1,000 documents, achieving 5.87 percent precision. The *Clustering – Weighted* strategy followed with 5.35 percent precision at the same recall level.

The *Random Sample* strategy achieved 7.30 percent precision at 75 percent recall using a seed set size of 2,000 documents. The *Stratified Keyword* strategy achieved 7.20 percent precision at the same recall level.

The *Random Sample* strategy, using a seed set size of 2,000 documents, achieved the highest precision of all the strategies at 75 percent recall. Further, predictive models created using seed set sizes of 2,000 documents achieved the six highest precision levels at 75 percent recall.

The Figure 2 illustrates the comparison of precision and recall levels on a curve for each strategy's seed set.

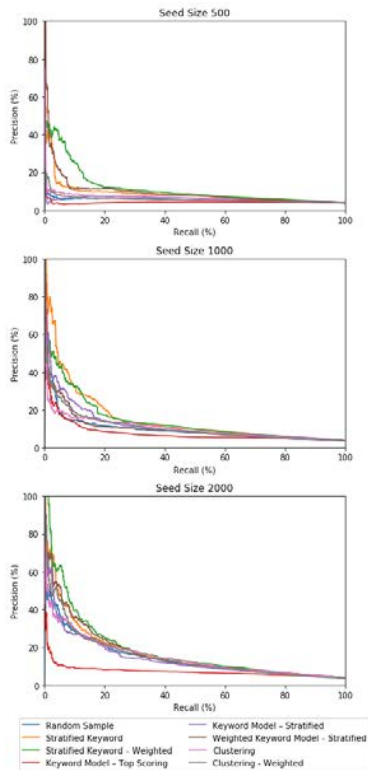

Figure 2. Project B - Precision & Recall by Strategy

## C. Project C

Tables 6A, 6B, and 6C illustrate the precision achieved at recall levels of 50 percent, 75 percent, and 90 percent using each strategy's seed set.

Table 6A: Project C - Seed Set 500 - Precision Rates at Recall Levels

| Seed Method | 90% | 75% | 50% |
|---|---|---|---|
| Clustering | 30.84% | 54.74% | 71.08% |
| Clustering – Weighted | 25.25% | 46.93% | 67.06% |
| Stratified Keyword – Weighted | 22.87% | 45.58% | 76.34% |
| Random Sample | 24.38% | 44.88% | 70.26% |
| Stratified Keyword | 23.06% | 43.91% | 76.09% |
| Weighted Keyword Model – Stratified | 22.32% | 43.79% | 72.96% |
| Keyword Model – Stratified | 20.98% | 40.33% | 69.97% |
| Keyword Model – Top Scoring | 21.39% | 25.55% | 31.25% |

Table 6B: Project C - Seed Set 1,000 - Precision Rates at Recall Levels

| Seed Method | 90% | 75% | 50% |
|---|---|---|---|
| Random Sample | 28.96% | 57.55% | 78.31% |
| Stratified Keyword – Weighted | 28.37% | 57.17% | 79.64% |
| Clustering | 34.29% | 55.87% | 76.15% |
| Stratified Keyword | 26.89% | 55.56% | 81.13% |
| Clustering – Weighted | 33.95% | 55.13% | 74.87% |
| Keyword Model – Stratified | 26.08% | 52.15% | 76.47% |
| Weighted Keyword Model – Stratified | 23.37% | 50.58% | 79.41% |
| Keyword Model – Top Scoring | 21.66% | 27.68% | 35.57% |

Table 6C: Project C - Seed Set 2,000 - Precision Rates at Recall Levels

| Seed Method | 90% | 75% | 50% |
|---|---|---|---|
| Clustering | 38.16% | 63.35% | 81.23% |
| Clustering – Weighted | 36.17% | 62.88% | 82.86% |
| Random Sample | 30.31% | 61.24% | 83.98% |
| Stratified Keyword | 30.40% | 59.90% | 87.75% |
| Stratified Keyword – Weighted | 28.99% | 54.66% | 84.95% |
| Keyword Model – Stratified | 25.57% | 54.50% | 83.07% |
| Weighted Keyword Model – Stratified | 24.52% | 51.85% | 83.10% |
| Keyword Model – Top Scoring | 19.09% | 25.70% | 37.22% |

The *Clustering* strategy achieved the highest precision at 75 percent recall using a seed set size of 500 documents, achieving 54.74 percent precision. The *Clustering - Weighted* strategy followed with 46.93 percent precision at the same recall level.

The *Random Sample* strategy achieved the highest precision at 75 percent recall using a seed set size of 1,000 documents, achieving 57.55 percent precision. The *Stratified Keyword – Weighted* followed with 57.17 percent precision at the same recall level.

The *Clustering* strategy achieved 63.35 percent precision at 75 percent recall using a seed set size of 2,000 documents. The *Clustering – Weighted* strategy followed closely with 62.88 percent precision at the same recall level.

The *Clustering* strategy, using a seed set size of 2,000 documents, achieved the highest precision of all the strategies at 75 percent recall. Further, predictive models created using seed set sizes of 2,000 documents achieved the four highest precision levels at 75 percent recall.

The Figure 3 illustrates the comparison of precision and recall levels on a curve for each strategy's seed set.

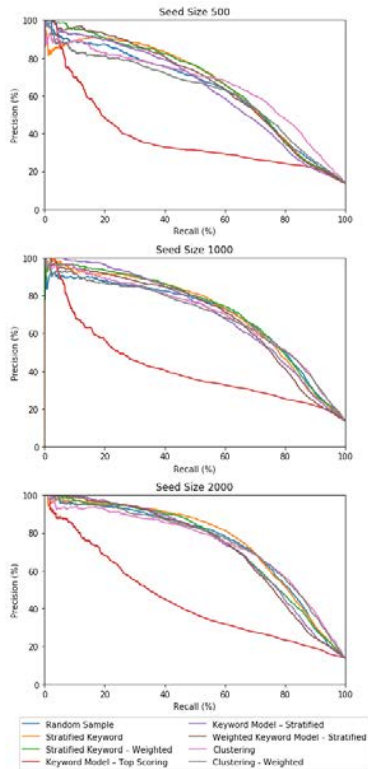

Figure 3. Project C - Precision & Recall by Strategy

## C. Project D

Tables 7A, 7B, and 7C illustrate the precision achieved at recall levels of 50 percent, 75 percent, and 90 percent using each strategy's seed set.

Table 7A: Project D - Seed Set 500 - Precision Rates at Recall Levels

| Seed Method | 90% | 75% | 50% |
|---|---|---|---|
| Random Sample | 73.60% | 88.58% | 95.33% |
| Clustering – Weighted | 70.00% | 88.52% | 94.79% |
| Clustering | 71.67% | 88.09% | 95.55% |
| Stratified Keyword | 65.90% | 84.22% | 95.09% |
| Stratified Keyword – Weighted | 65.61% | 82.39% | 93.05% |
| Keyword Model – Stratified | 62.63% | 82.05% | 94.07% |
| Weighted Keyword Model – Stratified | 61.04% | 82.04% | 94.20% |
| Keyword Model – Top Scoring | 41.46% | 44.17% | 50.80% |

Table 7B: Project D - Seed Set 1,000 - Precision Rates at Recall Levels

| Seed Method | 90% | 75% | 50% |
|---|---|---|---|
| Clustering | 76.58% | 91.35% | 96.21% |
| Clustering – Weighted | 76.75% | 91.12% | 96.14% |
| Random Sample | 74.08% | 89.86% | 96.24% |
| Weighted Keyword Model – Stratified | 66.28% | 86.54% | 95.86% |
| Keyword Model – Stratified | 64.90% | 86.12% | 96.35% |
| Stratified Keyword | 66.79% | 85.61% | 96.36% |
| Stratified Keyword – Weighted | 66.16% | 84.85% | 96.15% |
| Keyword Model – Top Scoring | 44.17% | 51.38% | 62.46% |

Table 7C: Project D - Seed Set 2,000 - Precision Rates at Recall Levels

| Seed Method | 90% | 75% | 50% |
|---|---|---|---|
| Clustering - Weighted | 78.04% | 91.58% | 96.53% |
| Random Sample | 78.05% | 91.50% | 96.53% |
| Clustering | 77.39% | 90.92% | 96.48% |
| Stratified Keyword | 69.77% | 88.72% | 96.57% |
| Stratified Keyword – Weighted | 64.08% | 84.77% | 96.69% |
| Keyword Model – Stratified | 60.05% | 82.58% | 95.55% |
| Weighted Keyword Model – Stratified | 60.67% | 81.89% | 95.48% |
| Keyword Model – Top Scoring | 48.34% | 58.73% | 75.44% |

The *Random Sample* strategy achieved the highest precision at 75 percent recall using a seed set size of 500 documents, achieving 88.58 percent precision. The *Clustering - Weighted* strategy followed with 88.52 percent precision at the same recall level.

The *Clustering* strategy achieved the highest precision at 75 percent recall using a seed set size of 1,000 documents, achieving 91.35 percent precision. The *Clustering – Weighted* strategy followed closely behind achieving 91.12 percent precision at the same recall level.

The *Clustering – Weighted* strategy achieved 91.58 percent precision at 75 percent recall using a seed set size of 2,000 documents. The *Random Sample* strategy's results were very similar, achieving 91.50 percent precision at the same recall level.

The *Clustering – Weighted* strategy, using a seed set size of 2,000 documents, achieved the highest precision all the strategies at 75 percent recall. Further, predictive models created using seed set sizes of 2,000 documents, achieved the top two precision levels at 75 percent recall.

The Figure 4 illustrates the comparison of precision and recall levels on a curve for each strategy's seed set.

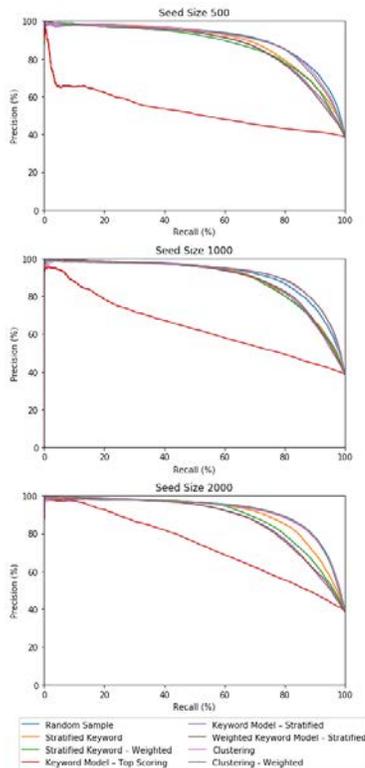

Figure 4. Project D - Precision & Recall by Strategy

## IV. CONCLUSIONS

Our study demonstrates that the strategy used to select a predictive coding model's seed set can have a significant impact on the future success of the process and perhaps even the legal case in general. However, legal teams rarely question if a seed set is selected using a random sample or a search term sample, or which strategy will provide the most effective results. The authors of this paper sought to identify selection strategies that minimize seed set size and increase recall and precision. This study reveals that different selection strategies have a measurable impact on the precision of a model at higher recall levels and that the strategy used to achieve high precision with smaller seeds sets often is not the same strategy that yields the highest precision when using larger seed sets.

This group of authors explored the performance of different selection strategies by conducting 24 experiments over four data sets, performing 96 experiments in total. The results demonstrated the following highlights:

- Seed sets of 2,000 documents repeatedly outperformed the seed sets with sizes of 500 and 1,000 documents. This makes sense considering more training exemplars should typically improve the performance of a predictive model.

- For Seed Sets of 2,000 documents, *Random Sample* and *Clustering* strategies performed in the top three strategies for each project regardless of richness of the data set.

- Table 8 summarizes the precision improvement when selecting the top performing 2,000 document seed set model for each project and compare it to the top performing 500 document seed set model for the same project.

Table 8: Precision Rate Difference from Seed Set size of 500 to 2,000

| Data Set | 90% Recall | 75% Recall | 50% Recall |
|---|---|---|---|
| Project A | 9.64% | 13.04% | 8.73% |
| Project B | 0.41% | 1.28% | 1.98% |
| Project C | 7.32% | 8.61% | 10.15% |
| Project D | 4.44% | 3.00% | 1.20% |

- *Keyword Model – Top Scoring* across all projects and seed set sizes was consistently the worst performing technique.

- For all projects, the top performing selection strategies for 500 document seed sets at 50 percent recall were: *Stratified Keyword – Weighted* and *Clustering*. Interestingly, these strategies' precision levels were, at most, 8.61 percent less precise than their related 2,000 document experiments. This is encouraging because legal teams endeavor to quickly locate relevant information to develop their understanding of the legal case's issues and to guide document review workflow. In practice, this means that legal teams can review fewer training documents and still create effective models to prioritize the initial phase of review.

- Table 9 summarizes a comparison of precision, at 50 percent recall, for seed set sizes of 500 and 2,000, for the best performing strategies for each project: *Stratified Keyword – Weighted* and *Clustering*.

Table 9 - *Stratified Keyword – Weighted* & *Clustering*: Precision at 50 Percent Recall

| Project | 2,000 Seed Set | 500 Seed Set | Differential |
|---|---|---|---|
| Project A: Stratified Keyword – Weighted | 81.11% | 74.32% | 6.79% |
| Project A: Stratified Keyword – Weighted | 12.16% | 8.56% | 3.60% |
| Project C: Stratified Keyword – Weighted | 84.95% | 76.34% | 8.61% |
| Project D: Clustering | 96.48% | 95.55% | 0.93% |

- In the data set with the lowest richness, Project B (3.63 percent), *Random Sample* performed the best when using a 2,000-document seed set but performed poorly when using a 500-document seed set. This is likely because there were substantially fewer positive training exemplars in the 500-document seed set when compared to the 2,000-document seed set. The *Stratified Keyword – Weighted* strategy achieved the highest precision using a 500-document seed set, improving precision by 23.80 percent when compared to *Random Sample*.

The study demonstrates that legal practitioners may achieve modest improvements over random sampling by employing well thought out techniques to ensure diversity of the seed set or to increase richness within the seed set. However, these results also serve as a caution: deviating from a random sampling technique can also result in measurably worse results, particularly where too much emphasis is placed on increasing richness in the seed set at the expense of ensuring diversity. In sum, this study provides new insights about predictive coding seed set strategies and counsel can use the results in this research to inform their seed set selection strategy decisions at the outset of a predictive coding review.